\documentclass[12pt, a4paper] {article}
\usepackage{cite}
\usepackage{amsfonts}
\usepackage{verbatim}
\usepackage{color}
\usepackage{amsmath}
\usepackage{graphicx}
\usepackage{bm}
\usepackage{amssymb}
\usepackage{latexsym}
\usepackage{mathrsfs}
\usepackage{textcomp}
\usepackage{ulem}
\begin{document}

\begin{center}

{\bf \large {(Anti-)BRST Symmetries in FLRW Model: \\Supervariable Approach}}

\vskip 2.5 cm

{\sf \bf{ Aradhya Shukla$^a$, Dharm Veer Singh$^a$ and R. Kumar$^b$}}\\
\vskip .1cm
{\textit {$^a$Department of Physics, Institute of Applied Sciences and Humanities, \\GLA 
University, Mathura--281406, Uttar Pradesh , India}}\\
{\textit {$^b$ Department of Physics, Siksha Bhavana, \\Visva-Bharati, Santiniketan, Bolpur--731235, West Bengal, India}}\\
\vskip 0.1 cm
{\sf {E-mails:  ashukla038@gmail.com; veerdsingh@gmail.com; raviphynuc@gmail.com}}
\end{center}
\vskip 2. cm

\noindent
\textbf {Abstract:} Friedmann--Lema$\hat{\text{i}}$tre--Robertson--Walker (FLRW) model, representing the isotropic and homogeneous Universe, has an inherent diffeomorphism (or reparametrization) invariance. For a given  local time-reparametrization symmetry transformation of
$(0+1)$-dimensional ($1D$) diffeomorphism  invariant FLRW model,  we derive the off-shell nilpotent and absolutely anticommuting (anti-)BRST symmetry transformations in a consistent manner. The supervariable approach to BRST formalism is used as a guiding tool in finding the proper (anti-)BRST transformations. An application of (super-)diffeomorphism invariance, in the context of superspace formalism, yields horizontality conditions. Moreover, the celebrated CF-type condition which is  {\it universal} for $1D$ diffeomorphism invariant theories, emerges as an off-shoot of the horizontality condition and provides the absolute anticommutativity of the (anti-)BRST transformations. In addition, this CF-type condition leads to the derivation of  coupled (but equivalent) (anti-)BRST invariant quantum Lagrangians for FLRW model. We also capture the geometrical interpretation of the (anti-)BRST transformations and their properties in terms of the Grassmannian translational generators and supervariables.



\newpage

\section{Introduction}
Quantization of gravitational theory is one of the most challenging  problems due to the high order of non-linearity. Several attempts such as weak field approximation~\cite{th,GTH,carl,fa,asp} and loop quantum gravity~\cite{asht,thie} have been made in this direction. The weak field approximation deals with the linearization of spacetime metric at perturbative level while
the loop quantum gravity is a non-perturbative and  background independent approach. Interestingly, both methods have revealed various important aspects of the Universe~\cite{mb,aa1,gw,ws,hco}.  It is to be noted that most of the theories of physical interest endowed with  constraints and reflect the gauge  and/or reparametrization (or diffeomorphism) invariance. Both gauge and reparametrization symmetries are local symmetries.  It is worth mentioning that in the Dirac's prescription for the classification scheme of constraints, the {\it first-class} constraints  generate the local gauge transformations~\cite{dirac1}.
On the other hand, the 
time-reparmetrization invariant theories  possess a {\it Hamiltonian} constraint which makes them  different from the usual gauge  theories. Since the reparametrization
invariance is one of the characteristic and distinguished  features of generally covariant system, it has attracted a great attention of the physicists to study and quantize those simple models having reparametrization (or diffeomorphism) invariance in the realm of quantum gravity and cosmological  models~\cite{ams,gonz,hal1}.

The generally covariant system of quantum gravity and quantum cosmology (having finite degrees of freedom) may be well described in terms of the Arnowitt--Deser--Misner (ADM) formulation~\cite{adm} using Friedmann--Lema$\hat{\text{i}}$tre--Robertson--Walker (FLRW) spacetime as a background  metric~\cite{FLRW1,FLRW2,FLRW3,FLRW4,FLRW5}. The latter one represents the homogeneous and isotropic  Universe.  One of the topical  issues in modern cosmology is related to the dark energy. The FLRW  metric lies at the heart of almost all the cosmological models and plays a central role, particularly, in the theoretical development of the dark energy models~\cite{das1,das2,das3}. These cosmological models, for instance, are described by the Einstein-Hilbert action coupled to a massless scalar field with cosmological constant in framework of FLRW spacetime metric. An intriguing feature of these models is that all possess a Hamiltonian constraint and, thus, the quantization can not be carried out in an usual way.

For the canonical quantization of constrained systems,  two different approaches involving the Lagrangian  and Hamiltonian of the system have been adapted.  The formulation of canonical quantization using Hamiltonian of the system has been established by Dirac in his seminal papers~\cite{dirac2,dirac3,dirac4}. In Dirac's canonical quantization of the reparametrization invariant theories, the  wavefunction representing the physical state of the system must be annihilated by the operator form of Hamiltonian constraint and, thus, consequently leads to the well-known  Wheeler-DeWitt equation~\cite{dwit}. 
In order to covariantly quantize a system, having local gauge symmetry,  the removal of redundant degrees of freedom requires to fix the gauge completely. The latter breaks the inherent local gauge symmetry of the system. For the restoration of symmetry, the  phase space must be extend. One of the ways to extend the phase space, by
adding extra degrees of freedom, is the Becchi--Rouet--Stora--Tyutin (BRST) formalism~\cite{brs1,brs2,brs3,brs4}.  In this formulation, the gauge-fixing and Faddeev-Popov (anti-)ghost terms are incorporated into the action to ensure the global supersymmetric type BRST invariance. 
In the BRST quantization,  the addition of Faddeev-Popov (anti-)ghost term does not change the physical content of the effective theory. 
It is well established that for a given   local gauge symmetry, in addition to global  BRST symmetry, we also have global anti-BRST symmetry.  Both BRST and anti-BRST symmetry transformations are nilpotent of order two and  absolutely anticommuting in nature~\cite{cf,oji}. Subsequently,  against the backdrop of BRST invariance, other  approaches  such as Batalin--Fradkin--Vilkovisky (BFV)~\cite{bfv1,bfv2} and  Batalin--Vilkovisky (BV) have been formulated~\cite{bv}.

One of the elegant and geometrically intuitive methods for obtaining the (anti-)BRST symmetry transformations corresponding to a local gauge symmetry is the Bonora-Tonin (BT) superfield formalism~\cite{bt1,bt2}. In the BT superfield approach, the (anti-)BRST transformations found their geometrical interpretation in terms of the Grassmannian translational generators. In this geometrical superfield formulation, one extends a general $D$-dimensional Minkowskian spacetime manifold parametrized by the bosonic coordinates $x^\mu$ (with $\mu = 0, 1, 2, ..., (D-1)$) to the $(D, 2)$-dimensional supermanifold ($Z^M = (x^\mu, \theta, \bar\theta))$ parametrized by two additional Grassmannian variables $\theta$ and $\bar\theta$. These Grassmannian variables, by construction, are nilpotent of order two ($\theta^2 = \bar\theta^2 = 0$) as well as anticommuting ($\theta \bar\theta + \bar\theta \theta = 0$). In addition, for example, the  ordinary  one-form connection $A^{(1)} = dx^\mu A_\mu$ and exterior derivative $d = dx^\mu \partial_\mu$ are generalised to super-one-form connection ${\cal A}^{(1)} = dZ^M {\cal A}_M = dx^\mu {\cal A}_\mu + d\theta\, \bar {\cal C} +   \bar d\theta\,  {\cal C}$ and super-exterior derivative  ${\widetilde d} = dx^\mu \, \partial_\mu + d \theta\,\partial_\theta + d \bar \theta\,\partial_{\bar\theta}$ onto $(D,2)$-dimensional supermanifold, respectively.
One of the important features of the superfield approach to BRST formalism is the natural emergence of Curci-Ferrari (CF) condition~\cite{cf}. This CF condition is required for the proof of  absolute anticommutativity of (anti-)BRST transformations as well as (anti-)BRST invariance of the  gauge-fixed coupled Lagrangian densities. In addition, CF condition also respects the BRST and anti-BRST transformations. In recent works~\cite{bm1,malik}, the superfield formalism has been applied to a class of local reparametrization (or diffeomorphism) invariant field-theoretic models and derived the (anti-)BRST symmetry transformations where the (super-)diffeomorphism invariance play a decisive role.

In our present endeavour, we consider a  generally covariant system of gravity in ADM formulation~\cite{adm}. One such system is the FLRW model of the homogeneous and isotropic  Universe~\cite{tps1,tps2}.  From the Lagrangian point of view,   
the FLRW model respects the time-reparametrization (i.e. $(0+1)$-dimensional diffeomorphism) symmetry transformation. The path integral quantization and the derivation of Wheeler-DeWitt equation for such time-reparametrization invariant models were carried out by Halliwell~\cite{hal2,hal3}. 
The quantum theory of FLRW model has been studied within the context of BV formalism, where the lapse function and scale factor are incorporated in the Lagrangian~\cite{tps1,tps2}. In FLRW model,  the lapse function and scale factor are known as ADM variables. 
Earlier, a few inconsistencies between Dirac and ADM approaches were pointed out~\cite{kiri} but in the context of BRST quantization these have been resolved by extending the phase space where the physical parameters and gauge degrees of freedom are  taken into account at equal footing~\cite{tps1,tps2}.
Furthermore, a suitable scalar product has been defined which ensures that the super-Hamiltonian for FLRW model vanishes. In addition,  a connection between the quantization of FLRW model by using the Lagrangian and Dirac's formulations of constrained dynamics has been established~\cite{cian}.

The purpose of our present investigation is three fold. \textit{First and foremost},  in earlier works~\cite{tps2,cian} only the BRST symmetry transformations  corresponding to the local 
time-reparametrization transformation were reported. As we know, for a given local symmetry, we have two independent supersymmetric type BRST and anti-BRST symmetry transformations. Thus, we  derive the complete sets of off-shell nilpotent {\it quantum} BRST and anti-BRST transformations. 
\textit{Second}, although, in  a recent work~\cite{bhup}, the anti-BRST transformations have been derived,
but a clear and consistent method for the derivation of (anti-)BRST transformations was missing; the (anti-)BRST transformations were not found to be anticommuting in nature. The latter is one of the key properties of the (anti-) BRST transformations. In our present work, we derive the anticommuting (anti-)BRST symmetry transformations in a neat and streamlined fashion.  Further, we show that the anticommuting property is satisfied due to the very  existence of CF-type condition. We also derive explicitly this CF-type condition within the framework of geometrical supervariable approach to BRST formalism. \textit{ Third}, in this work, by exploiting the basic tenets of BRST formalism, we construct the (anti-)BRST invariant coupled (but equivalent) Lagrangians for the FLRW model.

The layout of our present endeavour is organized as follows. In Section \ref{sec.2}, we discuss very briefly about the FLRW model and associated gauge and reparametrization symmetries. Our Section \ref{sec.3} deals with the derivation of off-shell nilpotent and anticommuting (anti-)BRST symmetry transformations corresponding to a local time-reparametrization transformation in the context of supervarable approach to BRST formalism. Section \ref{sec.4} is devoted to the discussion on the nilpotency and anticommutativity properties of the (anti-) BRST transformation as well as the validity of the CF-type condition. The (anti-)BRST invariance of the CF-type of condition is shown in this section, too.  By exploiting the basic tenets of BRST formalism, we derive the coupled (but equivalent) Lagrangians  in Section \ref{sec.5}. In Section \ref{sec.6}, we conclude the present work by summarizing our results and discuss the implication of our findings for the future investigation. 
Finally, in  Appendix A, we provide an explicit derivation of the CF-type of restriction.

\section{Gauge and Reparmetrization Symmetries of FLRW Model}
\label{sec.2}
Universe at large scale is described by the FLRW metric. This metric is based on the cosmological principle which minimally states that Universe is isotropic and homogeneous. In the  spherical coordinates $(t, r, \theta, \phi)$, the FLRW metric takes  the following form~\cite{tps1,tps2,cian}
\begin{eqnarray}
ds^2 = N^2 dt^2 + a^2 \bigg(\dfrac{dr^2 }{1-\kappa r^2} + r^2 d\theta^2 + r^2  sin^2 \theta d \phi^2 \bigg),
\label{1}
\end{eqnarray}
where  the lapse function $N(t)$ and scale factor $a(t)$  explicitly depend on the time-evolution  parameter $t$. It is to be noted that the parameter $\kappa = 1, 0, -1$ represents the closed, flat and open Universe,  respectively. In ADM formulation, the Lagrangian corresponding to FLRW metric in $(0+1)$-dimensional (1D) spacetime  can be written as~\cite{tps1,tps2}
\begin{equation}
L = -\frac{a\, {\dot a}^2}{2N} +\frac{\kappa}{2} N\, a.
\label{2}
\end{equation}
From the above Lagrangian, one  can readily  observe that the velocity $\dot N$ corresponding to the lapse function $N$  is missing. Thus, the Lagrangian has a  primary constraint 
\begin{eqnarray}
\Omega_1 = \pi_N \approx 0,
\label{3}
\end{eqnarray}
where $\Omega_1$ denotes the primary constraint on the theory and  $\pi_N = \partial L/\partial \dot N = 0$ defines the canonical momentum corresponding to the generalized variable $N$. Using the Legendre transformations, the canonical Hamiltonian $H_c$ of the system reads
\begin{eqnarray}
H_c = \dot a \, \pi_a + \dot N \, \pi_N  - L
= -\dfrac{N}{2} \left(\dfrac{\pi^2_a}{a} + \kappa a\right).
\label{4}
\end{eqnarray} 
In the above, $\pi_a = -a\, \dot a/N$ is the canonical momentum corresponding to the dynamical variable $a$. It can be easily checked that the canonical conjugate variables obey the following Poisson bracket algebra, namely;
\begin{eqnarray}
\big\{a, \;\pi_a \big\}_{PB} = 1, \qquad \big\{N, \;\pi_N \big\}_{PB} = 1.
\label{5}
\end{eqnarray}
and rest of the Poisson brackets vanish. In the above, the bracket $\{\;,\; \}_{PB}$ represents the Poisson bracket.

Now, for the manifestation of the physical constraints present in the theory, we introduce a first-order Lagrangian $(L_f)$ which is defined in the following fashion:
\begin{eqnarray}
L_f &=& \dot a \, \pi_a + \dot N \, \pi_N  - H_c = \dot a\, \pi_a + \dfrac{N}{2} \left(\dfrac{\pi^2_a}{a} + \kappa a\right).
\label{6}
\end{eqnarray}
It is clear from the above Lagrangian that the generalized coordinate $N$ behaves as an auxiliary (or Lagrange multiplier) variable. Thus, the corresponding momentum $\pi_N \equiv \Omega_1 \approx 0$  defines a primary constraint on the theory. Further, the coefficient of $N$ in the first-order Lagrangian $L_f$ enforces another constraint on the theory.  As we know that constraints remain intact with time, the persistence (i.e. $\dot \Omega_1   \approx 0$) of $\Omega_1$ leads to the secondary  constraint. Using the Euler-Lagrange equation of motion derived from $L_f$ w.r.t. $N$, we obtain the secondary constraint $\Omega_2$ as 
\begin{eqnarray}
\dfrac{d\Omega_1}{dt} = \dfrac{d}{dt}\left(\dfrac{\partial L_f}{\partial \dot N} \right) \approx 0  \Rightarrow \Omega_2=
\frac{1}{2}\Big(\frac{\pi^2_a}{a} + \kappa a \Big)\approx 0.
\label{7}
\end{eqnarray}
The above secondary constraint $\Omega_2$ can also be derive by exploiting the Hamilton's equation:
$\dot \Omega_1 = \big\{\Omega_1, H_c \big\}_{PB} \approx 0 \Rightarrow \Omega_2\approx 0$.
It can be readily checked that, the evolution of secondary constraint $\Omega_2$ does not yield any further constraint because $\{\Omega_2, \; H_c\}_{PB} = 0$. At this juncture, one can check that both $\Omega_1$ and $\Omega_2$ are first-class constraints (i.e. $\{\Omega_1, \; \Omega_2\}_{PB} = 0$) in the Dirac's classification scheme of constraints.

As a consequence of the above, the present model is a gauge-theoretic model endowed with first-class constraints. Moreover, these first-class constraints are the generators of the local gauge symmetry transformations. For the present FLRW model,  the gauge symmetry transformations generator is given as~\cite{ht1}
\begin{equation}
G = \pi_N \dot \Lambda - \frac{1}{2}\Big(\frac{\pi^2_a}{a} + \kappa a \Big) \Lambda,
\label{8}
\end{equation}
where $\Lambda =\Lambda(t)$ is a local gauge transformation parameter. Using the above generator, we obtain the following gauge transformations ($\delta_g$):
\begin{eqnarray}
&&\delta_g a = - \frac{\Lambda}{a} \pi_a, \qquad \delta_g \pi_a = \frac{\Lambda}{2} 
\Big(\kappa  - \frac{\pi^2_a}{a^2}  \Big), \qquad
\delta_g N = \dot \Lambda, \qquad \delta_g \pi_N = 0,
\label{9}
\end{eqnarray}
which leave the Lagrangian  (\ref{6})  quasi-invariant. In other words, Lagrangian transforms to a total time derivative as one can check
\begin{eqnarray}
\delta_g L_f =  \dfrac{d}{dt}\left[-\dfrac{\Lambda}{2} \left(\dfrac{\pi^2_a}{2} - \kappa a \right) \right].
\end{eqnarray} 
Thus,  the corresponding action integral remains invariant under the application of gauge transformations (\ref{9}).

Interestingly, we lay emphasis on the fact that, the existence of secondary constraint $\Omega_2 \approx 0$ enforces $H_c = 0$ on the constraint hypersurface defined by the first-class constraints. The vanishing of  canonical Hamiltonian $H_c = 0$ is the key signature of reparametrization invariance.  As a result, the above first-order Lagrangian remains  invariant under the  time-reparametrization (i.e. $(0+1)$-dimensional diffeomprphism) $t 
\rightarrow t' = f(t) = t - \xi(t)$.  
Here $f(t)$ is an arbitrary well-defined  function of time-evolution parameter $t$ and 
$\xi(t)$ is a local time-reparametrization transformation parameter.  Under the time-reparmetrization, the dynamical variables $a(t)$, $\pi_a(t)$ transform as scalar functions while $N(t)$ transforms like a density function. Mathematically, these can be expressed as 
\begin{eqnarray}
a(t) \to  a'(t') = a(t), \quad  \quad
\pi_a(t) \to \pi'_a(t') = \pi_a(t), \quad \quad
 N(t)\,dt \to  N'(t')\,dt' = N(t)\,dt.  
\end{eqnarray}
The infinitesimal version of the above transformations lead to the following time-reparametrization symmetry transformations $(\delta_r)$, respectively 
\begin{eqnarray}
\delta_r a =\xi\, \dot a, \qquad \quad 
\delta_r \pi_a =\xi \,\dot \pi_a, \qquad \delta_r N =\dot \xi \,N + \xi \,\dot N.
\label{12}
\end{eqnarray}
In the above, the subscript $r$ represents the local time-reparmetrization transformation ($\delta_r$). 
The first-order Lagrangian $L_f$ respects the above time-reparametrization symmetry transformations (\ref{12}) as one can check
\begin{eqnarray}
\delta_r L_f &=&  \frac{d}{dt}\Bigg[\xi \Bigg(\dot a \,\pi_a + 
\frac{N}{2}\Big(\frac{\pi^2_a}{a} - \kappa a\Big)\Bigg)\Bigg] 
= \frac{d}{dt} \Big(\xi \,L_f \Big).
\end{eqnarray} 
As a consequence, the action integral  remains invariant (i.e. $\delta_r S = \delta_r \int L_f dt =0 $) under the application of local time-reparametrization  transformations for the physically well-defined variables which vanish-off rapidly at $t\to \pm\infty$.

Before we wrap up this section, we make the following remark. At first glance, both the transformations in  Eq.~(\ref{9}) and Eq.~(\ref{12})  appear to be different as they originate from different roots. To be more precise, the gauge transformations (\ref{9}) are generated from the first-class constraints while the time-reparametrization transformations (\ref{12}) are obtained from  $(0+1)$-dimensional diffeomorphism invariance.  But, a close observation revels that both are  equivalent \textit{on-shell}. In fact, replacing $\Lambda \to \xi N $ together with the equation of motion $\dot \pi_a = \big\{\pi_a, H_c \big\}_{PB} = - (N/2)(\pi^2_a/a^2 - \kappa )$ in the  time-reparametrization transformations (\ref{12}), we recover the  gauge transformations in Eq.~(\ref{9}).

\section{Derivation of proper (anti-)BRST transformations: Supervariable approach}
\label{sec.3}
It is evident from the above discussion,  the first-order Lagrangian $L_f$ remains quasi-invariant under the time-reparametrization transformations $(\delta_r)$ corresponding to the $1D$ diffeomorphism i.e.  $t \rightarrow t' = f(t) = t-\xi(t)$.  In the superspace formulation, we generalize an ordinary $1D$
spacetime manifold parametrized by the time-evolution (bosonic) parameter $t$ to the $(1, 2)D$ supermanifold   parametrized by the superspace coordinates $T^M = (t, \theta, \bar \theta)$ where $(\theta, \bar \theta)$ are a pair of Grassmannian coordinates (with $\theta^2 = 0,\, \bar \theta^2 = 0,\, \theta\bar \theta + \bar \theta \theta = 0$). In addition $\theta$ and $\bar \theta$ carry ghost numbers $+1$ and $-1$, respectively. The generalization of an ordinary diffeomorphism $t \to t' = f(t)$ onto  $(1, 2)D$ supermanifold  is given by the super-diffeomorphism i.e. $T^M \to {\widetilde T}^M = ({\cal F}(T), \theta, \bar \theta)$ where ${\cal F}(T)$ is the  generalization of an ordinary function $f(t)$ onto $(1,2)D$ supermanifold\footnote{The Grassmannian coordinates ($\theta, \bar \theta$) and Faddeev-Popov ghost and anti-ghost ($C$ and $\bar C$) obey the following Hermiticity properties: $\theta^\dagger = \theta$, $\bar {\theta}^\dagger 
= - \bar \theta$, $C^\dagger = C$ and ${\bar C}^\dagger = - \bar C.$ The time-evolution parameter $t$ and local function $f(t)$ are assume to be real i.e. $t^\dagger = t$ and  $f^\dagger(t) = f(t)$}~\cite{bm1,malik}
\begin{eqnarray}
f(t) \to {\cal F}(T) =  {\cal F}(t, \theta, \bar \theta) &=& t- \theta \,\bar C(t) - \bar \theta \,C(t) + \theta \bar \theta\, h(t). 
\label{14}
\end{eqnarray}
In the above, $C, \bar C$ are the Grassmannian odd variables (with ghost numbers $+1$ and $-1$, respectively). The secondary variable $h$ and  time-evolution parameter $t$ are the Grassmannian even variables (with ghost numbers equal to zero). We shall, later on, identify that $C$ and $\bar C$ play the role of Faddeev-Popov ghost and anti-ghost variables of our BRST-invariant theory. It is to be noted that ${\cal F}(T)\big|_{\theta, \bar \theta = 0} = f(t)=t$.

\subsection{(Anti-)BRST transformations of dynamical variable $a(t)$ and its corresponding momentum  $\pi_a(t)$}
\label{subsec.3.1}

It is clear from our earlier Section \ref{sec.2} that the dynamical variables $a(t)$ and $\pi_a(t)$ obey the following one-dimensional diffeomorphism property
\begin{eqnarray}
&& a'(t') = a'\big(f(t)\big)  = a(t), \qquad \qquad
 \pi_a'(t') = \pi_a'\big(f(t)\big)  = \pi_a(t).
\label{15}
\end{eqnarray}  
As a consequence, we embed these scalar variables onto $(1, 2)D$ supermanifold. The generalization of the dynamamical variables onto $(1, 2)D$ supermanifold is given by the following super-expansions:
\begin{eqnarray}
a(t) \to {\cal A}(T) &=& {\cal A}(t, \theta, \bar \theta) = a(t) + \theta\, \bar R(t) + \bar \theta \,R(t) + \theta \bar \theta \,S(t), \nonumber\\
\pi_a(t) \to \Pi_{\cal A}(T) &=& \Pi_{\cal A} (t, \theta, \bar\theta) =  \pi_a(t) + \theta \, \bar 
P(t) + \bar \theta \, P(t) + \theta \bar \theta \,U(t),
\label{16}
\end{eqnarray}
where ${\cal A}(T) = {\cal A} (t, \theta, \bar\theta)$ and $\Pi_{\cal A}(T) = \Pi_{\cal A} (t, \theta, \bar\theta)$ are the supervariables corresponding to the dynamical variables $a(t)$ and $\pi_a(t)$, respectively. These dynamical variables carry ghost number zero.
In the above super-expansions,  $R$, $\bar R$, $P$, $\bar P$ (with ghost numbers equal to $ +1, -1, +1, -1,$ respectively) represent the secondary fermionic variables and $S$, $U$ (with ghost numbers zero) denote the secondary bosonic variables.  In the context  of  BT supervariable approach to BRST formalism, one can define the BRST $(s_b)$ and anti-BRST $(s_{ab})$ transformations in terms of the Grassmannian translational generators $\frac{\partial}{\partial \bar \theta},\; \frac{\partial}{\partial \theta}$, respectively. For the generic variable $\phi(t) \,(= a(t), \pi_a(t))$ and corresponding supervariable $\Phi(t, \theta, \bar \theta)$, we obtain
\begin{eqnarray}
s_b \phi(t) = \dfrac{\partial}{\partial_{\bar \theta}}\Phi(t, \theta, \bar \theta)\Bigg|_{\theta = 0}, \qquad \qquad
 s_{ab} \phi(t)= \dfrac{\partial}{\partial_{\theta}}\Phi(t, \theta, \bar \theta)\Bigg|_{\bar\theta = 0}, 
\end{eqnarray}    
where $\Phi(t, \theta, \bar \theta) = {\cal A} (t, \theta, \bar\theta),\; \Pi_{\cal A}(t, \theta, \bar\theta)$. In other words, the coefficients of $\bar \theta$ and $\theta$ in the super-expansions of the supervariables (cf. Eq. (16)) are nothing but the BRST $(s_b)$ and anti-BRST $(s_{ab})$ transformations, respectively. Therefore, in the operator form, we have the mappings  between the  (anti-)BRST transformations $(s_{(a)b})$ and Grassmannian  translational generators $\big(\frac{\partial}{\partial \theta},\; \frac{\partial}{\partial\bar \theta} \big)$, namely; $s_b \leftrightarrow \frac{\partial}{\partial \bar \theta} \Big|_{\theta = 0}$ and   $s_{ab} \leftrightarrow \frac{\partial}{\partial  \theta} \Big|_{\bar \theta = 0}$.

Under the application of super-diffeomorphism $(T^M \to {\widetilde T}^M = ({\cal F}(T), \theta, \bar \theta))$, the  supervariables ${\cal A}(t, \theta, \bar \theta)$ and 
$\Pi_{{\cal A}}(t, \theta, \bar \theta) $ transform as $\widetilde {\cal A} \big({\cal F}(T), \theta, \bar \theta)$ and $\widetilde {\Pi}_{\cal A} \big({\cal F}(T), \theta, \bar \theta)\big)$, respetively, in the following fashion:
\begin{eqnarray}
{\cal A}(t, \theta, \bar \theta) \to \widetilde {\cal A} \big({\cal F}(T), \theta, \bar \theta) \big)&=&  a \big({\cal F} (t, \theta, \bar\theta) \big) + \theta \, \bar 
R\big({\cal F} (t, \theta, \bar\theta)\big) \nonumber\\
&+& \bar \theta \, R\big({\cal F} (t, \theta, \bar\theta)\big) + \theta \bar \theta\, S\big({\cal F} (t, \theta, \bar\theta)\big), \nonumber\\
\Pi_{\cal A}(t, \theta, \bar \theta) \to \widetilde {\Pi}_{\cal A} \big({\cal F}(T), \theta, \bar \theta)\big) 
&=& \pi_a \big({\cal F}(t, \theta, \bar\theta)\big) - \theta\,  \bar P\big({\cal F}(t, \theta, \bar\theta)\big) \nonumber\\
 &-& \bar \theta \, P\big({\cal F} (t, \theta, \bar\theta) \big) + \theta \bar \theta \,U\big({\cal F} (t, \theta, \bar\theta)\big),
\label{18}
\end{eqnarray}
where the supervariables $a ({\cal F} \big(t, \theta, \bar\theta)\big), \;\, R \big({\cal F} (t,\theta, \bar\theta)\big), \; \,\bar R\big({\cal F} (t, \theta, \bar\theta)\big)$ and $S\big({\cal F} (t, \theta, \bar\theta)\big)$ are the  super-diffeomporphism generalization of the variables $a(t)$, $R(t)$, $\bar R(t)$  and $S(t)$, respectively, onto (1-2)-dimensional supermanifold  within the framework of supervariable approach to BRST formalism. The former can be further expressed, using Eq.~(\ref{14}), as
\begin{eqnarray}
a \big({\cal F} (t, \theta, \bar\theta)\big) &=& a \big(t- \theta \,\bar C- \bar\theta \,C + \theta \bar \theta \,h \big) \equiv  a - \theta \,\bar C \,\dot a - \bar \theta \, C \,\dot a + \theta \bar\theta \big(h\, \dot a - \bar C \,C \,\ddot a \big),\nonumber\\
R \big({\cal F} (t, \theta, \bar\theta)\big) &=& R \big(t- \theta \,\bar C- \bar\theta \,C + \theta \bar \theta \,h \big) \equiv  R - \theta\, \bar C \,\dot R - \bar\theta\, C\, \dot R 
+  \theta \bar\theta \big(h \,\dot R - \bar C\, C \,\ddot R \big),\nonumber\\
\bar R \big({\cal F} (t, \theta, \bar\theta)\big) &=& \bar R \big(t- \theta \,\bar C- \bar\theta\, C+ \theta \bar \theta \,h \big) \equiv  \bar R - \theta \,\bar C\, \dot {\bar R} - \bar\theta \, C \,
\dot {\bar R} +  \theta \bar\theta \big(h\, \dot {\bar R} - \bar C\, C\, \ddot {\bar R} \big),\nonumber\\
S \big({\cal F} (t, \theta, \bar\theta)\big) &=& S \big(t- \theta \,\bar C- \bar\theta\, C+ \theta \bar \theta \,h \big) \equiv  S - \theta \bar C \,\dot S - \bar\theta \,C \,\dot S +  \theta \bar\theta \big(h\, \dot S - \bar C \,C\, \ddot S \big). \qquad
\label{19}
\end{eqnarray}
Similarly, the supervarables $\pi_a \big({\cal F} (t, \theta, \bar\theta)\big)$, $P ({\cal F} \big(t, \theta, \bar\theta)\big)$, $\bar P\big({\cal F} (t, \theta, \bar\theta)\big)$ and $U\big({\cal F} (t, \theta, \bar\theta)\big)$ are the generalization of the  variables $\pi_a(t)$, $P(t)$, $\bar P(t)$ and $U(t)$, respectively, under the application of super-diffeomorphism onto (1, 2)D supermanifold. The super-expressions of these supervariables read
\begin{eqnarray}
\pi_a \big({\cal F} (t, \theta, \bar\theta)\big) &=& \pi_a \big(t- \theta \,\bar C- \bar\theta\, C + \theta \bar \theta \,h \big) \equiv  \pi_a - \theta \,\bar C\, \dot \pi_a - \bar\theta \, C \,\dot \pi_a + \theta \bar\theta \big(h\, \dot \pi_a - \bar C\, C \,\ddot \pi_a \big),\nonumber\\
P \big({\cal F} (t, \theta, \bar\theta)\big) &=& P \big(t- \theta \,\bar C- \bar\theta \,C + \theta \bar \theta\, h \big) \equiv  P - \theta \,\bar C \,\dot Q - \bar\theta \, C\, \dot P +  \theta \bar\theta \big(h \,\dot P - \bar C \,C \, \ddot P \big),\nonumber\\
{\bar P} \big({\cal F} (t, \theta, \bar\theta)\big) &=& \bar P \big(t- \theta \,\bar C- \bar\theta \,C + \theta \bar \theta \,h \big)  \equiv  \bar P - \theta \,\bar C \,\dot {\bar P} - \bar\theta \, C\, 
\dot {\bar P} +  \theta \bar\theta \big(h \,\dot {\bar P} - \bar C\, C\, \ddot {\bar P} \big),\nonumber\\
U \big({\cal F} (t, \theta, \bar\theta)\big) &=& U \big(t- \theta \,\bar C- \bar\theta\, C + \theta \bar \theta \,h \big) \equiv  U  - \theta \,\bar C \,\dot  U - \bar\theta \, C\, \dot U +  \theta \bar\theta \big(h\, \dot U - \bar C\, C\, \ddot U\big). \qquad
\end{eqnarray}
In the above, dot over the variables denotes differentiation w.r.t. the time-evolution parameter $t$.

In view of the (super-)diffeomorphism properties (cf. Eq. (\ref{15}) and Eq.~(\ref{18})) of the  dynamical variables $a(t)$ and $\pi_a(t)$, we define the following horizontallity conditions (HCs):
 \begin{eqnarray}
\widetilde {\cal A}\big({\cal F}(T), \theta, \bar \theta\big)&=&
\widetilde {\cal A}\big({\cal F}(t, \theta, \bar \theta),  \theta, \bar \theta\big) = a(t), \nonumber\\
\widetilde{\Pi}_{\cal A}\big({\cal F}(T), \theta, \bar \theta\big) &=&
\widetilde{\Pi}_{\cal A}\big({\cal F}(t, \theta, \bar \theta),  \theta, \bar \theta\big)=\pi_a(t).
\label{21}
\end{eqnarray}
These  HCs imply that the scalar variables/functions under the application of (super-)diffeomorphism remain unchanged.  In other words, the scalar variables/fields
do not transform at all under any kind of transformations such as spacetime translation, internal transformation, supersymmetric transformation, etc.

Using Eqs. (\ref{19})--(\ref{21}), we obtain the following interesting relationships among the basic and secondary variables:
\begin{eqnarray}
&& R = C \,\dot a, \qquad \quad \bar R = \bar C\, \dot a, \quad \qquad P = C \,{\dot \pi}_a, \qquad \quad \bar P = \bar C \,{\dot \pi}_a, \nonumber\\
&& S= -\Big(\big(h + \dot {\bar C} \,C + \bar C \,\dot C \big) \dot a + \bar C\, C\, \ddot a \Big), \qquad 
U= -\Big(\big(h + \dot {\bar C} \,C + \bar C \,\dot C \big) {\dot \pi}_a + \bar C \,C\, {\ddot \pi}_a \Big). \qquad \quad
\label{22}
\end{eqnarray}
Substituting these relationships in Eq.~(\ref{16}), we obtain the desired expressions of the supervariables, namely;
\begin{eqnarray}
{\cal A}(t, \theta, \bar\theta) &=& 
a + \theta \big(\bar C \,\dot a \big) + \bar\theta \big(C\, \dot a\big) + \theta \bar\theta \Big(
\big(B - {\bar C}\, \dot  C \big) \dot a + {\bar C}\, C\, {\ddot a} \Big) 
\nonumber\\
&\equiv& a + \theta \big[s_{ab} \,a\big] + \bar\theta \big[s_{b}\,a\big] + \theta \bar\theta 
\big[s_b s_{ab}\, a\big],\nonumber\\
\Pi_{\cal A}(t, \theta, \bar\theta) &=& \pi_a 
+ \theta \big(\bar C \,{\dot \pi}_a \big) + \bar\theta \big(C \,\dot {\dot \pi}_a \big) + \theta \bar\theta \Big(\big(B -  {\bar C} \,\dot C\big) {\dot \pi}_a - {\bar C} \,C {\ddot  \Pi}_a\Big) \nonumber\\
&\equiv& \pi_a + \theta \big[s_{ab}\,  \pi_a\big] + \bar\theta \big[s_{b}\, \pi_a\big] + \theta 
\bar\theta \big[s_b s_{ab}\, \pi_a\big], 
\label{23}
\end{eqnarray}
where we have purposely chosen $B = -(h + \dot {\bar C} \,C)$. We shall see later on, that $B$  will play the role of Nakanishi-Lautrup auxiliary variable of the BRST-invariant theory. It is to be noted that the terms in  square brackets represent the (anti-)BRST transformations for the generic variable $\phi = a,\, \pi_a$.

\subsection{(Anti-)BRST transformations of lapse function $N(t)$ and associated (anti-)ghost variables $(\bar C)C$ }
\label{subsec.3.2}
For the derivation of proper (anti-)BRST transformations of the lapse function $N(t)$, we recall the property of one-dimensional differomphism i.e. $N(t)$ transforms like a density function under time-reparametrization (i.e.  $N'(t') dt' = N(t)\, dt$). In the language of differential form, the density function $N(t)\, dt = N^{(1)} $ defines a one-form onto $(0+1)$-dimensional manifold. Thus, the corresponding super-one-form ${\cal N}^{(1)} = {\cal N}_M (T) \,d T^M $ onto $(1, 2)D$ supemanifold is given as
\begin{eqnarray}
{\cal N}_M (T)\, d T^M &=& {\cal N}_t(t, \theta, \bar \theta)\, dt 
+ {\cal N}_\theta (t, \theta, \bar \theta)\,d\theta + {\cal N}_{\bar\theta}(t, \theta, \bar \theta)\,d\bar\theta,
\label{24}
\end{eqnarray}
with following supermultiples, namely; 
\begin{eqnarray}
{\cal N}_t(t, \theta, \bar\theta) &=&  N(t) + \theta\, \bar \lambda(t) +  \bar\theta \,
\lambda(t)  + \theta \bar\theta\, b(t), \nonumber\\
{\cal N}_\theta (t, \theta, \bar\theta) &=&  \zeta(t) + \theta\, \bar \alpha(t)  +  
\bar\theta \,\alpha(t)  + \theta \bar\theta\, \sigma(t),\nonumber\\
 {\cal N}_{\bar\theta} (t, \theta, \bar\theta) &=&  \omega(t) + \theta \,\bar 
\beta(t) +  \bar\theta \,\beta(t)  + \theta \bar\theta\, \rho(t),
\label{25}
\end{eqnarray}
where,  $b, \alpha, \bar \alpha, \beta, \bar \beta$ are secondary bosonic variables and $\lambda, \bar \lambda, \sigma, \rho$ are the secondary fermionic  variables. The secondary variables $\zeta$ and $\omega$ are also fermionic in nature.  We shall determine these secondary variables in terms of the basic variables of the (anti-)BRST invariant theory. The dynamical variables $N(t)$, $\zeta(t)$ and $\omega(t)$ carry ghost numbers equal to $0,\;-1$ and $+1$, respectively.  The ghost numbers of the remaining secondary variables being (bosonic)fermionic can be read/understood from the above super-expansions (cf. Eq.~(\ref25)).

In the faith of (super-)diffeormorphism invariance, we  define another HC as~\cite{bm1} 
\begin{eqnarray}
\widetilde{\cal N}_M(\widetilde T)\, {\widetilde 
d}\,{\widetilde T}^M = N(t) \,dt,
\label{26}
\end{eqnarray}
where
${\widetilde d} = dt\, \partial/\partial t + d\theta\, \partial/\partial \theta + d\bar\theta\, 
\partial/\partial {\bar\theta}$ defines the super-exterior derivative onto $(1, 2)D$ supemanifold corresponding to an ordinary exterior derivative $d = dt\,\partial/\partial t$ defined onto $(0+1)D$ manifold. In the above, the quantity ${\widetilde 
d}\,{\widetilde T}^M $ can be written in matrix form as
\begin{eqnarray}
\widetilde d\, \widetilde T^M &=& \widetilde d \big({\cal F}(T), \theta, \bar\theta \big) \nonumber\\
&=& \Big(dt \dfrac{\partial}{\partial t} + d\theta \dfrac{\partial}{\partial \theta} + d\bar\theta 
\dfrac{\partial}{\partial {\bar\theta}}\Big)\big(t-\theta \,\bar C - \bar\theta\, C +\theta \bar \theta\, h, \theta, \bar \theta\big)\nonumber\\
 &=& \Big(\big(1 - \theta\, \dot{\bar C} - \bar\theta \,\dot C + 
\theta\bar\theta\, \dot h \big) dt + \big(\bar C - \bar\theta\, h\big)d\theta + \big( C + \theta \,h\big) d\bar\theta, \,d\theta,\, d\bar\theta\Big). 
\label{27}
\end{eqnarray}
Using Eqs. (\ref{24}), (\ref{25}) and (\ref{27}), we explicitly compute the l.h.s. of the  HC (cf. Eq.~(\ref{26})) as
\begin{eqnarray}
\widetilde {\cal N}_M (\widetilde T)\, \widetilde d \widetilde T^M &=& \Big[N + \theta \big(\bar \lambda - \bar C \,\dot N - \dot {\bar C} \,N \big) + \bar \theta \big(\lambda - C \,\dot N - \dot C\, N \big) \nonumber\\
&+& \theta \bar \theta \Big(b + \dfrac{d}{dt} \big(h\,N + \bar C\, \lambda - C\, \bar \lambda - \bar C\, C\, \dot N \big) \Big)\Big]dt \nonumber\\
&+& \Big[\zeta -\bar C \,N + \theta \big(\bar \alpha - \bar C \,\dot \zeta +  \bar C \,\bar \lambda\big) + \bar \theta \big(\alpha + h\, N -  C \,\dot \zeta +\bar C \,\lambda - \bar C \,C \,\dot N \big) \nonumber\\
&+& \theta \bar \theta \big(\sigma + h \,\dot \zeta - h\, \bar \lambda + \bar C\, \dot \alpha - b\, \bar C - C\, \dot{\bar \alpha} + \bar C\, C \,\dot {\bar \lambda}- {\bar C} \,C\, \ddot \zeta \big)\Big] d\theta \nonumber\\
&+& \Big[\omega + C\, N + \theta \big(\bar \beta +h \,N - C\, \bar \lambda - \bar C \,\dot \omega - \bar C\, C\, \dot N \big) 
+ \bar \theta \big(\beta - C\, \lambda -  C \,\dot \omega \big) \nonumber\\
&+& \theta \bar \theta \big(\rho + h\, \dot \omega +h \,\lambda +
b\, C - C \,\dot {\bar \beta} + \bar C \,\dot \beta-  \bar C \,C \dot \lambda -\bar C\, C\, \ddot \omega \big)\Big] d \bar\theta. 
\label{28}
\end{eqnarray}
Using the above Eq.~(\ref{28}) and  HC in Eq.~(\ref{26}),  the coefficients of the differential $dt$, $d \theta$ and $d \bar \theta$ yield the following relationships amongst the basic dynamical and secondary variables as: 
\begin{eqnarray}
&& \lambda = C \,\dot N  + \dot C \,N, \qquad \bar \lambda = \bar C\, \dot N  + \dot {\bar C} \,N, \qquad \zeta = \bar C \,N, \qquad \omega = - C\,N, \qquad b = \dot \alpha, \nonumber\\
&& \alpha = N\,B - \bar C \big(\dot C\, N + C \,\dot N \big),  \qquad \bar \beta = - N \,\bar B - \big(\dot  {\bar C}\, N  + \bar C \,\dot N \big) C, \qquad \alpha = \bar\beta, \nonumber\\
&& \bar \alpha = 0,  \qquad \rho = 0, \qquad \beta  = 0,  \qquad  \sigma =0,
\label{29}
\end{eqnarray}
where, for the sake of brevity, we have chosen $B = -(h + \dot {\bar C} \,C)$ and $\bar B = (h +  \bar C \, \dot C)$ which 
play the role of Nakanishi-Lautrup type auxiliary variables. After making these choices for the Nakanishi-Lautrup type auxiliary variables, the super-expansions of the supervariables will take the simplest form (cf. Eqs.~(\ref{30}) and (\ref{32})). As a result, we obtain the  standard and desired (anti-)BRST transformations (see, Section \ref{sec.4} below).

Substituting the above relationships (cf. Eq.~(\ref{29})) into the expressions of supervariables listed in Eq.~(\ref{25}), we obtain
\begin{eqnarray}
{\cal N}_t(t, \theta, \bar\theta) &=&  N + \theta \big(\bar C \,\dot N + \dot{\bar C}\, N \big) +  \bar\theta \big(C \,\dot N + {\dot C}\, N \big) + \theta \bar\theta 
\Big[\frac{d}{dt} \Big(N\,B -\bar C \big(\dot C\,N + C\,\dot N \big)\Big)\Big] \nonumber\\
&\equiv& N + \theta \big[s_{ab} N \big] +  \bar\theta \big[s_b N \big] + \theta \bar\theta 
\big[s_b s_{ab} N\big],\nonumber\\
{\cal N}_\theta (t, \theta, \bar\theta) &=& \bar C\, N + \bar \theta \Big(N\,B - \bar C \big(\dot C \,N + C\, \dot N \big)\Big) \nonumber\\
 &\equiv&    \bar C\, N + \theta \big[s_{ab} (N\, \bar C)\big] + \bar\theta \big[s_b (N\bar 
C) \big] + \theta \bar\theta \big[s_b s_{ab} (N\,\bar C)\big],\nonumber\\
{\cal N}_{\bar\theta} (t, \theta, \bar\theta) &=&  -C\,N + \theta \Big(-N\,{\bar B} - \big(\dot{\bar C} \,N + \bar C \, \dot N\big) C\Big) \nonumber\\
 &\equiv& -C\,N + \theta \big[s_{ab} \big(-C\,N \big)\big] + \bar \theta \big[s_b \big(-C\,N \big)\big] + \theta \bar\theta \big[s_b s_{ab} \big(-C\,N \big)\big].
\label{30}
\end{eqnarray}
The above super-expansions, however, reveal the (anti-)BRST transformations of $N$ and composite variables $\bar CN$ and $-CN$, the (anti-)BRST transformations of ghost and anti-ghost variables as well as corresponding supervariables are still unknown. For this purpose, we first compute the inverse ${\cal N}^{-1}_t (t, \theta, \bar\theta) $ of ${\cal N}_t (t, \theta, \bar\theta)$ as\footnote{For any given bosonic  supervariable $\Phi(t, \theta, \bar \theta) = \phi(t) + \theta \,\bar\eta(t) + \bar \theta \,\eta(t) + \theta \bar \theta \,\gamma(t)$ corresponding to a non-zero variable $\phi(t)$, the  inverse of $\Phi(t, \theta, \bar \theta)$ is given as:
$\Phi^{-1}(t, \theta, \bar \theta) = \dfrac{1}{\phi}\Big[1- \theta\,\dfrac{1}{\phi}\, \bar \eta - \bar \theta\,\dfrac{1}{\phi}\, \eta -\theta \bar \theta\, \dfrac{1}{\phi}\Big(\gamma + \dfrac{2}{\phi} \bar \eta \eta \Big) \Big].$
} 
\begin{eqnarray}
{\cal N}^{-1}_t (t, \theta, \bar\theta) &=& \dfrac{1}{N}\bigg[1 -\theta \bigg(\dfrac{\bar C\, \dot N + \dot {\bar C}\, N}{N} \bigg)
- \bar \theta \bigg(\dfrac{C \,\dot N + \dot C\, N }{N}\bigg) \nonumber\\
&-& \theta \bar \theta  \bigg(\dfrac{d}{dt}\Big(N\,B - \bar C \big(C\, \dot N + \dot C\, N \big) \Big) + \dfrac{2}{N} \big(\bar C \,\dot N -\dot {\bar C} \,N \big) \big(C\, \dot N + \dot C \,N \big) \bigg)\bigg].
\label{31} \quad \;
\end{eqnarray}

Now, with the help of  Eqs. (\ref{30}) and (\ref{31}), we   define the supervariables $\bar {\cal C}(t, \theta, \bar\theta)$ and  ${\cal C}(t, \theta, \bar\theta)$ corresponding to the ordinary (anti-)ghost variables  $\bar C$ and  $C$, respectively
\begin{eqnarray}
\bar {\cal C}(t, \theta, \bar\theta)= {\cal N}_\theta (t, \theta, \bar\theta) \, {\cal N}^{-1}_t (t, \theta, \bar\theta) 
&=& \bar C + \theta \big(\bar C \,\dot {\bar C} \big) + \bar\theta \,B - \theta\bar\theta\big({\dot B}\, \bar C - B\, \dot{\bar C} \big) \nonumber\\
&\equiv & \bar C + \theta \big[s_{ab}\, \bar C \big] 
+ \bar \theta \big[s_b\, \bar C \big] + \theta \bar \theta
 \big[s_b s_{ab}\, \bar C \big], \nonumber\\
{\cal C}(t, \theta, \bar\theta)= - {\cal N}_{\bar \theta} (t, \theta, \bar\theta)\,{\cal N}^{-1}_t (t, \theta, \bar\theta) 
&=& C + \theta \,\bar B + \bar\theta \big(C\,\dot C \big) + \theta\bar\theta\big(\dot{\bar B}\, C - \bar B \,\dot C\big), \nonumber\\
&\equiv &  C + \theta \big[s_{ab}\,  C \big] 
+ \bar \theta \big[s_b\, \bar C \big] + \theta \bar \theta
 \big[s_b s_{ab}\,  C \big].
 \label{32}
\end{eqnarray}
Thus, we obtain the (anti-)BRST transformations of the (anti-)ghost variables $(\bar C)C$ as the coefficients of $(\theta) \bar\theta$, respectively. It is worth noting that the lapse function $N(t)$ must be different from zero. Because, in the limit $N(t) = 0$, the supervariables defined in Eqs.~(\ref{30})--(\ref{32}) are not
well-defined. In other words, if $N(t) = 0$, the horizontality condition given in Eq.~ (\ref{26}) does not exist and the whole program of the supervariable approach to BRST formalism spoils completely. Further, the Lagrangian $L$ in Eq.~({\ref 2}) is not well-defined when $N(t) = 0$.  
 
Before we wrap up this subsection, a couple of points are in order. 
{\it First,} we point out that the algebraic relationships in the last line of Eq.   (\ref{29}) yield the  celebrated CF-type condition\footnote{For the explicit derivation of CF-type condition, see Appendix A. }
\begin{eqnarray}
B + \bar B + \big(\dot{\bar C}\, C - \bar C \,\dot C\big) = 0.
\label{33}
\end{eqnarray}
This CF-type condition, as we shall see later,  plays a crucial role in the absolute anticommutativity of the BRST and anti-BRST transformations (see, Eq.~(\ref{37}) below). And, it will also play an important in the derivation of coupled Lagrangians which respect both BRST and anti-BRST transformations 
(see, Section~\ref{sec.5} below). {\it Second}, the (anti-)BRST transformations of the Nakanishi-Lautrup type auxiliary variables $B$ and ${\bar B}$ have not yet been determined because we do not know the expressions  of supervariables corresponding to the Nakanishi-Laurtup type auxiliary variables. These variables appeared in the super-expansions of $(\bar {\cal C}){\cal C}$ as the secondary variables (cf. Eq.~(\ref{32})) and they are  also related to the secondary variable $h$ defined in the super-expansion of  ${\cal F}(T) = {\cal F}(t, \theta, \bar\theta)$  (cf. Eq.~(\ref{14})). Although, the super-expansions for $B$ and $\bar B$ can be written, if the (anti-)BRST transformations of $B$ and $\bar B$ are known. We shall determine the (anti-)BRST transformations for $B$ and $\bar B$ using the key properties (i.e. nilpotency and absolute anticommutativity properties) of (anti-)BRST transformations in our upcoming Section \ref{sec.4}.

\section{(Anti-)BRST transformations: Nilpotency,   anticommutativity check and (anti-)BRST invariance of Curci-Ferrari type of condition}
\label{sec.4}
From the above Eqs. (\ref{23}), (\ref{30}) and (\ref{32}), one can read-off the  BRST transformation ($s_b$) and anti-BRST  transformation ($s_{ab}$) for all the variables. These symmetry transformations are listed as
\begin{eqnarray}
&& s_b N = \dot C \,N + C \,\dot N, \qquad s_b a = C \,\dot a, \qquad s_b \pi_a = C \,{\dot \Pi}_a,  \qquad s_b C = C \,\dot C, \nonumber\\
&& s_b \bar C = B, \qquad s_b \bar B = \dot{\bar B} \,C - {\bar B} \,\dot C, \qquad s_b B = 0, \label{34}\\
&&\nonumber\\
&& s_{ab} N = \dot {\bar C}\, N + {\bar C} \,\dot N, \qquad s_{ab} a = \bar C\, \dot a, \qquad s_{ab} \pi_a = \bar C \,{\dot \pi}_a, \qquad
 s_{ab} {\bar C} = {\bar C}\, \dot {\bar C}, \nonumber\\
&& s_{ab} C = \bar B, \qquad s_{ab}  B = \dot B \,{\bar C} -  B\, \dot {\bar C},\qquad s_b {\bar B} = 0.
\label{35}
\end{eqnarray}
The above BRST and anti-BRST transformations, by construction of supervariable approach to BRST formalism, obey the two key properties, namely;  the off-shell nilpotency 
of order two (i.e. $s^2_b =0$ and $s^2_{ab} = 0$) and absolute anticommutivity property ($s_b\,s_{ab}+ s_{ab}\,s_b = 0$). The nilpotency property reveals the fact that the (anti-)BRST transformations are fermionic in nature whereas the absolute anticommutativity property implies that the (anti-)BRST transformations are linearly independent of each other.

We lay emphasize on the fact that the (anti-)BRST transformations of Nakanishi-Lautrup type auxiliary variables $B$ abd $\bar B$ have been derived from the off-shell nilpotency 
of order two and absolute anticommutivity property of the (anti-)BRST transformations $s_{(a)b}$. To be more precise, the BRST transformations $s_b B = 0,\; s_{ab} \bar B =0$ and $s_b\bar B = \dot {\bar B} C- \bar B \dot C,\; s_{ab} B= \dot B \bar C - B \dot {\bar C}$ can be obtained by the requirements of
nilpotency and absolute anticommutativity of the (anti-)BRST transformations, respectively. As one can readily check that the following 
\begin{eqnarray}
s_b^2 \bar C = 0 &\Rightarrow & s_b B = 0, \nonumber\\
s_{ab}^2  C = 0 &\Rightarrow & s_{ab} \bar B = 0, \nonumber\\
\big(s_b\,s_{ab}+ s_{ab}\,s_b \big) C = 0 &\Rightarrow & s_b \bar  B =\dot {\bar B} \,C - \bar B\, \dot  C, \nonumber\\
\big(s_b\,s_{ab}+ s_{ab}\,s_b \big)\bar C = 0 &\Rightarrow & s_{ab} B =\dot B\, \bar C - B\, \dot {\bar C},
\label{36}
\end{eqnarray}
lead to the correct (anti-)BRST transformations of the Nakanishi-Lautrup type auxiliary variables $B$ ans $\bar B$. We further point out that  the Nakanishi-Lautrup type auxiliary variables $(\bar B) B$ are not the basic dynamical variables of the (anti-) BRST invariant theory. In fact, they are required for the off-shell nilpotency of the (anti-)BRST transformations $s_{(a)b}$. It is straightforward to check that above fermionic (anti-)BRST transformations $(s_{(a)b})$ are off-shell nilpotent of order two (i.e. $s^2_{(a)b} = 0$).


As far as the absolute anticommutativity property of the (anti-)BRST transformations is concerned, it is to be noted that 
the absolute anticommutativity property of (anti-)BRST transformations for the dynamical variables $a, \pi_a, N$ is satisfied only on the constrained submanifold defined by the CF-type condition (\ref{33}). For the sake of brevity, the anticommutativity property for these variables yields
\begin{eqnarray}
\big(s_b \,s_{ab} + s_{ab}\, s_b \big) a &=& \big[B  + \bar B + \big(\dot {\bar C}\, C - \bar C \,\dot C\big)\big]\dot a =0,\nonumber\\
\big(s_b \,s_{ab} + s_{ab} \,s_b\big) \pi_a &=& \big[B  + \bar B + \big(\dot {\bar C}\, C - \bar C\, \dot C\big)\big]{\dot \pi}_a =0,\nonumber\\
\big(s_b \,s_{ab} + s_{ab} \,s_b\big)N &=& \dfrac{d}{dt}\Big[\Big(B  + \bar B + \big(\dot {\bar C} \,C - \bar C\, \dot C\big)\Big)N\Big] = 0. \label{37}
\end{eqnarray}
In the above, the l.h.s. turns out to be zero because of the validity of CF-type condition (\ref{33}). For rest of the variables $C, \bar C, B, \bar B$, the anticommutativity property is trivially satisfied. As a consequence, the 
(anti-)BRST transformatins are linearly independent symmetry transformations provided the entire theory is to be considered on the constrained submanifold defined by CF-type condition i.e.  $B  + \bar B + \big(\dot {\bar C}\, C - \bar C \,\dot C\big)$. It is a physical condition/restriction on our (anti-)BRST invariant theory.

Here, we emphasize that the (anti-)BRST transformations of (anti-)ghost variables $(\bar C)C$ can also be determined without taking any recourse of the expressions of (anti-)ghost supervariables (\ref{32}). We can compute them 
directly from the expressions of the  supervariables ${\cal N}_{\bar \theta} (t, \theta, \bar \theta)$ and  ${\cal N}_{\theta} (t, \theta, \bar \theta)$ (cf. Eq.~(\ref{30})) and the known (anti-)BRST transformations of $N$ (cf. Eqs.~(\ref{34}) and (\ref{35})). For instance, we note that the following are true:
\begin{eqnarray}
&&\dfrac{\partial}{\partial_{\bar \theta}} {\cal N}_{\bar\theta} (t, \theta, \bar \theta)\bigg|_{\theta = 0} =  s_b \big(-C\, N \big) = 0 \;\Rightarrow \; \big(s_b\,  C \big)N - C \big(s_b\, N\big)  =0,\nonumber\\
&& \dfrac{\partial}{\partial_{\theta}} {\cal N}_\theta (t, \theta, \bar \theta)\bigg|_{\bar \theta = 0} =  s_{ab} \big(\bar C\, N \big) = 0 \;\Rightarrow \;\big(s_{ab}\,  \bar C \big)N - \bar C \big(s_{ab}\, N\big)  =0. 
\label{38}
\end{eqnarray}  
For the given values of $s_b N$ and $s_{ab} N$,  we obtain the desired transformations: $s_b C = C \,\dot C$ and $s_{ab} \bar C = \bar C \,\dot {\bar C}$.  Furthermore, one can also reproduce these transformations using the nilpotency property of (anti-)BRST transformations, as one can   immediately check  that $s^2_b\, a = 0 \Rightarrow s_b C = C\, \dot C$  and $s^2_{ab}\, a = 0 \Rightarrow s_{ab} \bar C = \bar C \,\dot{\bar C}$.

Similarly, the transformations $s_b\bar C = B$ and $s_{ab}C= B$ can be obtained from the super-expansion ${\cal N}_\theta (t, \theta, \bar \theta)$  and  ${\cal N}_{\bar \theta} (t, \theta, \bar \theta)$, respectively, in a straightforward manner. It is clear from the expressions of supervariables (\ref{30}) that 
\begin{eqnarray}
\dfrac{\partial}{\partial \bar \theta} {\cal N}_\theta (t, \theta, \bar \theta)\bigg|_{\theta = 0} = s_b (\bar C\, N) &=& \big(s_b \bar C \big)N - \bar C \big(s_b N\big) \nonumber\\
 &=& N\,B - \bar C\big(\dot C\, N + C\, \dot N \big)\nonumber\\
&=& N \,B - \bar C \big(s_b N \big), \nonumber\\
&& \nonumber\\
\dfrac{\partial}{\partial \theta} {\cal N}_{\bar \theta} (t, \theta, \bar \theta)\bigg|_{\bar\theta = 0} =  s_{ab} (-C \,N) &=&  
- \big(s_{ab} C\big)N + C \big(s_{ab} N\big)  \nonumber\\
&=& - N\,\bar B - \big(\dot {\bar C}\, N + \bar C \,\dot N \big) C \nonumber\\
&=& - N \bar B  +C \big(s_{ab} N\big),
\end{eqnarray}
yield $s_b \bar C = B$  and $s_{ab} C = \bar B$, respectively, as listed above. 
Therefore, with the help of (anti-)BRST transformations (cf. Eqs. (\ref{34}) and (\ref{35})), we can easily construct the (anti-)ghost supervariables $\bar {\cal C}(t, \theta, \bar\theta)$, ${\cal C}(t, \theta, \bar\theta)$ and the result will turn out to be exactly same as quoted in Eq. (\ref{32}).

As we have shown above, the (anti-)BRST transformations are linearly independent symmetry transformations due to the validity of CF-type of condition (cf. Eq.~ (\ref{33})); this condition must respect the (anti-)BRST transformations. Therefore,  we here explicitly show the  (anti-)BRST invariance of CF-type of condition $B + \bar B + \big(\dot {\bar C}\, C- \bar C \, \dot C \big) = 0$. To prove this assertion,  we  note that 
the action of BRST and anti-BRST transformations on the l.h.s. of CF-type restriction yields 
\begin{eqnarray}
s_b \big[B + \bar B + \big(\dot {\bar C}\, C- \bar C \, \dot C \big) \big] &=& - \big[B + \bar B + \dot {\bar C}\, C \big]\, \dot C + \big[\dot B + \dot {\bar B}  - \bar C \, \ddot C \big] \, C \nonumber\\
&=& -\big[B + \bar B + \big(\dot {\bar C}\, C- \bar C \, \dot C \big)\big]\, \dot C +\big[\dot B+ \dot {\bar B} + \big(\ddot {\bar C}\, C - \bar C \, {\ddot C}\big)\big] \, C \nonumber\\
&=& -\big[B + \bar B + \big(\dot {\bar C}\, C- \bar C \, \dot C \big)\big]\, \dot C \nonumber\\
&+&\dfrac{d}{dt}\big[B+ {\bar B} +  \big(\dot {\bar C}\, C- \bar C \, \dot C \big)\big] \, C, 
\label{5a}\\
&&\nonumber\\
s_{ab} \big[B + \bar B + \big(\dot {\bar C}\, C- \bar C \, \dot C\big)\big] &=& -\big[B + \bar B - \bar C \, \dot C\big]\,\dot {\bar C} + \big[\dot B + \dot {\bar B} + {\ddot {\bar C}} \, C\big] \, \bar C \nonumber\\
&=& -\big[B + \bar B + \big(\dot {\bar C}\, C- \bar C \, \dot C\big)\big]\, \dot {\bar C} + \big[\dot B+ \dot {\bar B} + \big(\ddot {\bar C}\, C - \bar C \, {\ddot C}\big)\big]\,\bar C \nonumber\\
&=& -\big[B + \bar B + \big(\dot {\bar C}\, C- \bar C \, \dot C\big)\big]\, \dot {\bar C}  \nonumber\\
&+&\dfrac{d}{dt}\big[B+ \bar B + \big(\dot {\bar C}\, C - \bar C \, \dot C\big)\big] \, \bar C. 
\label{5b} 
\end{eqnarray}
In the above, we have used the fact that Faddeev-Popov (anti-)ghost variables are fermionic in nature (i.e.  $ \bar C^2 = 0,\; C^2 = 0, \;  \dot {\bar C}^2 = 0$ and $\dot C^2 =0$) and $\dot B + \dot {\bar B} + \big(\ddot {\bar C}\, C - \bar C \, {\ddot C} \big) = \dfrac{d}{dt}\big[B + \bar B + \big(\dot {\bar C}\, C- \bar C \, \dot C \big) \big]$.

It is now clear from the r.h.s. of the above Eqs.~(\ref{5a}) and (\ref{5b}) that
\begin{eqnarray}
s_b[B + \bar B + (\dot {\bar C}\, C - \bar C \, \dot C)] =0, \qquad
s_{ab}[B + \bar B + (\dot {\bar C}\, C - \bar C \, \dot C)] = 0,
\label{5c}
\end{eqnarray}
are true only on the submanifold defined by the CF-type of condition (cf. Eq.~(\ref{33})). Thus, the submanifold defined by the CF-type of condition remains invariant under the (anti-) BRST symmetry transformations.

Before we wrap up this section, we point out that the nilpotency and absolute anticommutativity properties of BRST and anti-BRST transformations can also be described in terms of  the Grassmannian derivatives. For any given generic supervariable $\Phi(t, \theta, \bar \theta)$ corresponding to an ordinary variable $\phi(t)$, the nilpotency and absolute anticommutativity properties of the (anti-)BRST transformations in terms of Grassmannian translational generators  read 
\begin{eqnarray}
s^2_b \, \phi(t) = 0 &\Leftrightarrow & \dfrac{\partial}{\partial \bar \theta}\,\dfrac{\partial}{\partial \bar \theta}\, \Phi(t, \theta, \bar \theta) =0, \nonumber\\
s^2_{ab} \, \phi(t) = 0 &\Leftrightarrow & \dfrac{\partial}{\partial \theta}\,\dfrac{\partial}{\partial \theta}\, \Phi(t, \theta, \bar \theta) =0, \nonumber\\
\Big(s_b\, s_{ab} + s_{ab}\, s_b \Big) \phi(t) &\Leftrightarrow & \bigg(\dfrac{\partial}{\partial \bar \theta}\,\dfrac{\partial}{\partial \theta} + \dfrac{\partial}{\partial \bar \theta}\,\dfrac{\partial}{\partial \bar \theta} \bigg)\Phi(t, \theta, \bar \theta) =0, 
\label{40}
\end{eqnarray}
where, the generic variables $\phi(t) = a, \pi_a, N, C, B, \bar B, \bar C$  are ordinary variables and their corresponding supervariables are defined as $\Phi(t, \theta, \bar \theta)$  (see, Eqs. (\ref{23}), (\ref{30}), (\ref{32}) and (\ref{54}) below). In the above, we have use nilpotency  $\big(\text{i.e.}\; \frac{\partial^2}{\partial \theta^2} =  \frac{\partial^2}{\partial \bar \theta^2} = 0 \big)$ and anticommutativity 
$\big(\text{i.e.}\; \frac{\partial}{\partial \theta} \frac{\partial}{\partial \bar \theta} + \frac{\partial}{\partial \bar \theta} \frac{\partial}{\partial  \theta} = 0\big)$  properties of the Grassmannian derivatives.

\section{(Anti-)BRST invariant coupled (but equivalent) Lagrangians}
\label{sec.5}
In order to construct the quantum theory of FLRW model, it is indispensable to fix the gauge degrees of freedom completely by imposing a suitable gauge-fixing condition. For this purpose, we choose the differential gauge-fixing condition~\cite{tps1,tps2}
\begin{eqnarray}
\dot N -\dot \chi(a) = 0,
\label{41}
\end{eqnarray} 
where, $\chi(a)$ is an arbitrary function of scale factor $a$.
This gauge condition (\ref{41}) may be incorporated into the action/Lagrangian with a Lagrange multiplier variable.  Although,
the gauge-fixing term breaks the \textit{ local} reparametrization invariance, the addition of Faddeev-Popov (anti-)ghost terms ensure the \textit{ global quantum} gauge (i.e. (anti-)BRST) invariance. 

Using the basic principles of BRST formalism, the gauge-fixed  BRST and anti-BRST invariant coupled Lagrangians $L_{(b)}$ and $L_{(ab)}$, respectively, can be written as~\cite{tps1,tps2}
\begin{eqnarray}
L_{(b)} &=& L_f + s_b \Big[\bar C\Big(\dot N - \dot \chi(a) \Big)\Big] \nonumber\\
&\equiv & L_f + B \Big(\dot N - \dot \chi(a) \Big) + \Big(\dot N - \dot\chi(a) \Big) \dot{\bar C}\, C+ N \,\dot{\bar C}\, \dot C, 
\label{42}
\end{eqnarray}
\begin{eqnarray}
L_{(ab)} &=& L_f - s_{ab} \Big[C\Big(\dot N - \dot \chi(a) \Big)\Big] \nonumber\\
&\equiv & L_f - \bar B \Big(\dot N - \dot \chi(a) \Big) + \Big(\dot N - \dot \chi(a) \Big) {\bar C} \,\dot C  + N\,\dot{\bar C} \,\dot C, 
\label{43}
\end{eqnarray}
modulo total time derivative. Here, the subscripts $(a), \, (ab)$ on the Lagrangians denote the BRST and anti-BRST invariant Lagrangians, respectively. The equations of motion derived from the above Lagrangians w.r.t. the auxiliary variables $B$ and $\bar B$ enforce the differential gauge-fixing condition $\dot N -\dot \chi(a) = 0$ (cf. Eq. (\ref{41})).  It is straightforward to check that both $L_{(b)}$ and $L_{(ab)}$ are coupled Lagrangians because the Nakanishi-Laurtup type auxiliary variables $B$ and $\bar B$ are related through CF-tpye condition (\ref{33}). Again,  both Lagrangians differ by an algebraic relation, as one can  readily check that 
\begin{eqnarray}
L_{(b)} - L_{(ab)} = \big(\dot N - \dot \chi(a)\big) \Big[B + \bar B + \big(\dot {\bar C} \,C - \bar C \,\dot C \big) \Big]. \;
\label{44}
\end{eqnarray}  
It is to be noted that r.h.s. is zero due the very existence  of CF-type condition: $B + \bar B +\dot {\bar C}\, C - \bar C \,\dot C = 0$. As a result, both Lagrangians  are equivalent $(i.e. L_{(b)} \equiv L_{(ab)} )$, too, on the constrained hypersurface defined by CF-type condition.

The Lagrangians $L_{(b)}$ and $L_{(ab)}$ respect both BRST and  anti-BRST symmetry transformations, respectively. Mathematically, this statement can be corroborated as 
\begin{eqnarray}
s_b\, L_{(b)} &=& \dfrac{d}{dt}\,\Big[\Big(\dot a \,\pi_a + \dfrac{N}{2} \Big(\dfrac{\pi^2_a}{a} + \kappa a \Big) \Big) C + B \big(\dot N - \dot \chi(a)\big) C + B \, N \dot C \Big],
\label{45}
\end{eqnarray}
\begin{eqnarray}
s_{ab}\, L_{(ab)} &=& \dfrac{d}{dt}\,\Big[\Big(\dot a \,\pi_a + \dfrac{N}{2} \Big(\dfrac{\pi^2_a}{a} + \kappa a \Big) \Big) \bar C 
- \bar B \,  \big(\dot N - \dot \chi(a)\big) \bar C - \bar B \, N \dot {\bar C} \Big].
\label{46}
\end{eqnarray}
Thus, the corresponding actions remain invariant for the physically well-defined variables which vanish-off rapidly at $t = \pm \infty$. As mentioned above, both Lagrangian are equivalent on the constraint hypersurface defined by Eq. (33), the Lagrangians $L_{(b)}$ and $L_{(ab)}$ must respect the anti-BRST and BRST transformations, respectively on the constraint hypersurface (\ref{33}). The application of $s_{ab}$ and $s_b$ on $L_{(b)}$ and $L_{(ab)}$, respectively yield
\begin{eqnarray}
s_{ab}\, L_{(b)} &=&  
\dfrac{d}{dt}\,\Big[\Big(\dot a \,\pi_a + \dfrac{N}{2} \Big(\dfrac{\pi^2_a}{a} + \kappa a \Big) \Big) \bar C + \big(B + \dot {\bar C} \, C\big) \big(\dot N - \dot \chi(a) \big) \bar C + \big(B - \bar C \,\dot C\big) \, N \dot {\bar C} \Big] \nonumber\\
&-&\Big[B + \bar B + \big(\dot {\bar C}\, C - \bar C \, \dot C \big) \Big] \big(\dot N - \dot \chi(a)\big)\dot {\bar C} \nonumber\\
&-& \Big[\dfrac{d}{dt}\Big(B + \bar B + \big(\dot {\bar C}\, C - \bar C \,\dot C\big)  \Big) \Big] N\, \dot {\bar C},
\label{47}
\end{eqnarray}
\begin{eqnarray}
s_{b}\, L_{(ab)} &=&  
\dfrac{d}{dt}\,\Big[\Big(\dot a \,\pi_a + \dfrac{N}{2} \Big(\dfrac{\pi^2_a}{a} + \kappa a \Big) \Big)\,C - \big(\bar B - {\bar C} \, \dot C\big) \big(\dot N - \dot \chi(a) \big) C - \big(\bar B + \dot {\bar C} \, C\big) \, N \dot C \Big] \nonumber\\
&+&\Big[B + \bar B + \big(\dot {\bar C}\, C - \bar C \, \dot C \big) \Big] \big(\dot N - \dot \chi(a)\big)\dot C \nonumber\\
&+& \Big[\dfrac{d}{dt}\Big(B + \bar B + \big(\dot {\bar C}\, C - \bar C \,\dot C\big)  \Big) \Big] N\, \dot  C,
\label{48}
\end{eqnarray}
total time derivatives  plus extra terms. These extra terms become zero due to the CF-type condition (\ref{33}). Thus,  the (anti-)BRST transformations $s_{(a)b}$ are the symmetries of both Lagrangians $L_{(a)b}$ on the constraint hypersurface defined by CF-type condition (\ref{33}).  Therefore, we can write 
\begin{eqnarray}
s_b L_{(b)} &\equiv& s_b L_{(ab)}\big|_{B + \bar B + (\dot {\bar C}\, C - \bar C \, \dot C) = 0}, \nonumber\\
 s_{ab} L_{(ab)} &\equiv &s_{ab} L_{(b)}\big|_{B + \bar B + (\dot {\bar C}\, C - \bar C \, \dot C) = 0}, 
\label{49}
\end{eqnarray}
which  show the equivalence of coupled Lagrangians on the constraint hypersurface (\ref{33}).

The above (anti-)BRST invariance can also be captured in terms of the supervariables and Grassmannian derivatives. To do this, we first generalize the first-order Lagrangian $L_f$ to super-Lagrangian ${\cal L}_f$ onto $(1,2)D$ supermanifold. The latter is expressed in terms of supervariables as
\begin{eqnarray}
L_f \to {\cal L}_f &=& \dot {\cal A}(t, \theta, \bar \theta)\, \Pi_{\cal A} (t, \theta, \bar \theta) + \dfrac{1}{2}\, {\cal N}_t(t, \theta, \bar \theta)\Big(\Pi^2_{\cal A}(t, \theta, \bar \theta)\,{\cal A}^{-1}(t, \theta, \bar \theta) + \kappa {\cal A}(t, \theta, \bar \theta) \Big) \nonumber\\
&=& L_f + \theta \dfrac{d}{dt} \Big[\bar C \,L_f \Big]  + \bar \theta \dfrac{d}{dt} \Big[C \,L_f \Big] + \theta\bar\theta \dfrac{d}{dt}\Big[B\, L_f - \bar C \,\dfrac{d}{dt} \Big({C \, L_f} \Big)\Big],
\label{50}
\end{eqnarray}
where ${\cal A}^{-1}(t, \theta, \bar \theta)$ is the inverse of  ${\cal A}(t, \theta, \bar \theta)$. It is evident form  Eq. (\ref{50}), the (anti-)BRST invariance of first-order Lagrangian can be expressed in terms of Grassmannian translational generators $\partial/\partial \theta,\; \partial/\partial {\bar \theta}$ and supervariables as
\begin{eqnarray}
\dfrac{\partial}{\partial \bar \theta} {\cal L}_f \bigg|_{\theta =0} = s_b \,L_f = \dfrac{d}{dt}\Big(C\, L_f\big), \qquad \qquad
\dfrac{\partial}{\partial \theta} {\cal L}_f \bigg|_{\bar \theta =0}  = s_{ab} \,L_f = \dfrac{d}{dt}\Big(\bar C \,L_f\big).
\label{51}
\end{eqnarray}
Similarly, the (anti-)BRST invariant Lagrangians can be expressed in terms of supervariables (cf. Eqs.~(\ref{23}), (\ref{30})--(\ref{32}))   in the following fashion:
\begin{eqnarray}
 {\cal L}_{(b)} &=& {\cal L}_f + {\cal B}(t, \theta, \bar \theta) \Big[\dot {\cal N}_t(t, \theta, \bar \theta)- \dot {\cal X}\big({\cal A}(t, \theta, \bar \theta)\big) \Big] \nonumber\\
 &+& \Big[\dot {\cal N}_t(t, \theta, \bar \theta)- \dot {\cal X}\big({\cal A}(t, \theta, \bar \theta)\big) \Big]
{\dot {\bar {\cal C}}}
(t, \theta, \bar \theta)\,{\cal C}
(t, \theta, \bar \theta) + {\cal N}_t(t, \theta, \bar \theta) \, {\dot {\bar {\cal C}}}
(t, \theta, \bar \theta) \,{\dot {\cal C}}
(t, \theta, \bar \theta),\label{52}\\
&&\nonumber\\
 {\cal L}_{(ab)} &=& {\cal L}_f - \bar {\cal B}(t, \theta, \bar \theta) \Big[\dot {\cal N}_t(t, \theta, \bar \theta)- \dot {\cal X}\big({\cal A}(t, \theta, \bar \theta)\big) \Big] \nonumber\\
 &+& \Big[\dot {\cal N}_t(t, \theta, \bar \theta)- \dot {\cal X}\big({\cal A}(t, \theta, \bar \theta)\big) \Big]\, {\bar {\cal C}}
(t, \theta, \bar \theta)\,\dot {\cal C}
(t, \theta, \bar \theta) +  {\cal N}_t(t, \theta, \bar \theta)\,{\dot {\bar {\cal C}}}(t, \theta, \bar \theta) \,{\dot {\cal C}}
(t, \theta, \bar \theta), \qquad
\label{53}
\end{eqnarray}
where  the supervariable ${\cal X}({\cal A}(t, \theta, \bar\theta))$ defined onto $(1,2)$-dimensional supermanifold is the generalization of an ordinary function $\chi(a)$. The former is given by the following expression
\begin{eqnarray}
{\cal X}({\cal A}(t, \theta, \bar\theta))&=&\chi(a) + \theta \,\bar C\, \dot \chi(a) + \bar \theta\, C \,\dot \chi(a)  \nonumber\\
&+& \theta \bar \theta \Big[\Big(B + \dot{\bar C} \,C\Big) \dot \chi(a) - \dfrac{d}{dt} \Big(\bar C\, C\, \dot \chi(a)   \Big)\Big], 
\label{54} 
\end{eqnarray}
and, the supervariables ${\cal B}(t, \theta, \bar \theta)$ and $\bar{\cal B}(t, \theta, \bar \theta)$ corresponding to 
the ordinary auxiliary variables $B$ and $\bar B$ read
\begin{eqnarray}
{\cal B}(t, \theta, \bar \theta)= B +  \theta \big(\dot B \,\bar C -  B \,\dot{\bar  C}\big), \nonumber\\
\bar{\cal B}(t, \theta, \bar \theta)= \bar B  + \bar\theta \big(\dot{\bar B}\, C - \bar B\, \dot C \big). 
\label{55}
\end{eqnarray}
Exploiting Eqs. (\ref{52})--(\ref{55}), we obtain
\begin{eqnarray}
\dfrac{\partial}{\partial \bar \theta} {\cal L}_{(b)} \bigg|_{\theta =0} = s_b \,L_{(b)} &=& \dfrac{d}{dt}\,\Big[\Big(\dot a \,\pi_a + \dfrac{N}{2} \Big(\dfrac{\pi^2_a}{a} + \kappa a \Big) \Big) C + B \big(\dot N - \dot \chi(a)\big) C + B \, N\, \dot C \Big],\quad\\
&&\nonumber\\
\dfrac{\partial}{\partial \theta} {\cal L}_{(ab)} \bigg|_{\bar \theta =0} = s_{ab} \,L_{(ab)} &=& \dfrac{d}{dt}\,\Big[\Big(\dot a \,\pi_a + \dfrac{N}{2} \Big(\dfrac{\pi^2_a}{a} + \kappa a \Big) \Big) \bar C - \bar B \,  \big(\dot N - \dot \chi(a)\big) \bar C - \bar B \, N \,\dot {\bar C} \Big], \qquad \; \\
\dfrac{\partial}{\partial \theta} {\cal L}_{(b)} \bigg|_{\bar \theta =0} = s_{ab} \,L_{(b)} &=&  \dfrac{d}{dt}\,\Big[\Big(\dot a \,\pi_a + \dfrac{N}{2} \Big(\dfrac{\pi^2_a}{a} + \kappa a \Big) \Big) \bar C + \big(B + \dot {\bar C} \, C\big) \big(\dot N - \dot \chi(a) \big) \bar C \nonumber\\
&+& \big(B - \bar C \,\dot C\big)  N\, \dot {\bar C} \Big] - \Big[B + \bar B + \big(\dot {\bar C}\, C - \bar C \, \dot C \big) \Big] \big(\dot N - \dot \chi(a)\big)\dot {\bar C} \nonumber\\
&-& \Big[\dfrac{d}{dt}\Big(B + \bar B + \big(\dot {\bar C}\, C - \bar C \,\dot C\big)  \Big) \Big] N\, \dot {\bar C},\\ 
\dfrac{\partial}{\partial \bar \theta} {\cal L}_{(ab)} \bigg|_{\theta =0} = s_b \,L_{(ab)} 
&=&  \dfrac{d}{dt}\,\Big[\Big(\dot a \,\pi_a + \dfrac{N}{2} \Big(\dfrac{\pi^2_a}{a} + \kappa a \Big) \Big) C - \big(\bar B - {\bar C} \, \dot C\big) \big(\dot N - \dot \chi(a) \big) C \nonumber\\
&-& \big(\bar B + \dot {\bar C} \, C\big) \, N\, \dot C \Big]  + \Big[B + \bar B + \big(\dot {\bar C}\, C - \bar C \, \dot C \big) \Big] \big(\dot N - \dot \chi(a)\big)\dot C \nonumber\\
&+& \Big[\dfrac{d}{dt}\Big(B + \bar B + \big(\dot {\bar C}\, C - \bar C \,\dot C\big)  \Big) \Big] N\, \dot  C. 
\end{eqnarray}
These equations imply the (anti-)BRST invariance of the coupled Lagrangians  within the framework of supervariable approach to BRST formalism (provided the whole theory is  considered on the  constrained hypersurface defined by CF-type condition $B + \bar B + \big(\dot {\bar C}\, C - \bar C \, \dot C \big)=0$.
Further, it is to be noted that the following mathematical expressions
\begin{eqnarray}
\dfrac{\partial}{\partial \bar \theta} {\cal L}_{(b)} \bigg|_{\theta =0} &\equiv & \dfrac{\partial}{\partial \bar \theta} {\cal L}_{(ab)} \bigg|_{\theta =0;\; B+\bar B+ \big(\dot {\bar C}\, C - \bar C \, \dot C \big) =0 }, \nonumber\\
\dfrac{\partial}{\partial  \theta} {\cal L}_{(ab)} \bigg|_{\bar \theta =0} &\equiv & \dfrac{\partial}{\partial \bar \theta} {\cal L}_{(b)} \bigg|_{\bar \theta =0;\; B+\bar B+ \big(\dot {\bar C}\, C - \bar C \, \dot C \big) =0},
\end{eqnarray}
show the equivalence of (super-)Lagrangians (i.e. $ {\cal L}_{(b)} \equiv {\cal L}_{(ab)} \Leftrightarrow  L_{(b)} \equiv L_{(ab)}$) due to the validity of CF-type condition (\ref{33}).

\section{Conclusions}
\label{sec.6}
In our present endeavour, we have studied the FLRW model of homogeneous and isotropic Universe within the framework of supervariable approach to BRST formalism. We have derived the off-shell nilpotent and absolutely anticommuting (anti-)BRST transformations corresponding to the local time-reparametrization transformation by exploiting the BT supervariable approach where, the property of (super-)diffeomorphism invariance is taken into account. Interestingly, the CF condition, necessary for the proof of absolute anticommutativity of the 
(anti-)BRST transformations,  emerges naturally in this approach. Moreover, using the basic principles of BRST formalism, we have explicitly obtained the  gauge-fixed (anti-)BRST invariant Lagrangians $L_{(b)}$ and $ L_{(ab)}$.  Both Lagrangians are coupled (but equivalent) because of the existence of CF-type condition (\ref{33}).
Further, a close observation reveals that both Lagrangians as well as (anti-)BRST transformations are symmetric if we interchange the role of $(\bar C)C$ and $(\bar B)B$. To be more precise, if we replace $C \to + \bar C$, $\bar C \to - C$, $B \to -\bar B$ one can go from $L_{(b)} \to L_{(ab)}$ and from $s_b \to s_{ab}$.

We point out that, in earlier work~\cite{bhup}, the complete sets of (anti-)BRST transformations for FLRW model have been obtained using the (anti-)BRST invariant restrictions. The impositions of (anti-)BRST restrictions to derive the (anti-)BRST transformations might not be the good idea and proper way because it seems obvious and trivial in some sense.  Further, The anti-BRST transformation is an independent symmetry transformation, it must anticommute with BRST transformation.  However, in Ref.~\cite{bhup}, the (anti-)BRST transformations were found to be non-anticommuting in nature.
In present investigation, we have exploited the more elegant and cultivated method to obtain the proper (anti-)BRST symmetry transformations in context of BT supervariable approach to BRST formalism where, the property of (super-)diffeomorphism invariance has been incorporated in a fruitful manner. In fact, in our derivation of the (anti-)BRST  transformations, we  have utilized  the fact that under the time-reparmetrization $t \to  t' = f(t) = t- \xi (t) $ the scale factor $a$ and associated momentum $\pi_a$ transform as scalar functions while the gauge variable (i.e. lapse function) $N$ as density function. The generalization of these facts together with super-diffeomorphism $T^M \to {\widetilde T}^M = ({\cal F}(T), \theta, \bar \theta)$ onto $(1, 2)D$ supermanifold  manifestly yields the appropriate HCs (cf. Eqs. (\ref{21}) and  (\ref{26})).

The application of HCs in Eq. (\ref{21}) produces the off-shell nilpotent (anti-) BRST transformations of the dynamical variables $a$ and $\pi_a$. 
It is to be noted that the density function $N(t)dt$  is equivalent to the one-form $N^{(1)} = N(t)dt$ in (0+1)-dimension (1D) of spacetime and remains unchanged under the ordinary diffeomorphism  (or time-reparmetrization) transformation and hence under the super-diffeomorphism. 
This yields another horizonatality (or soul-flatness) condition (\ref{26}) for the lapse function $N$. Thus, the HC in Eq. (\ref{26}) leads to the proper (anti-) BRST transformations of the gauge variable $N$ and corresponding (anti-)ghost variables $(\bar C)C$. One of the novel outcomes of this horizontality condition is the  existence of CF-type condition $B +\bar B + \big(\dot {\bar C}\, C - \bar C \dot C \big) =0$ (cf. Eq. (\ref{33})).  The CF-type condition is found to be BRST as well as anti-BRST invariant on the constrained submanifold defined itself by the CF-type restriction. It is a physical restriction on our (anti-)BRST invariant theory.  We lay emphasize that it is an {\it universal} condition for the 
time-reparametrization (or $1D$ differomorphism) invariant theories~\cite{malik}.     This condition plays an important role in providing the absolute anticommutativity of the (anti-)BRST transformations for the dynamical variables $a, \pi_a$ and $N$ (cf. Eq.~(\ref{37})) provided the  complete theory is considered on a submanifold defined by the CF-type restriction. We lay stress here that the (anti-)BRST transformations of Nakanishi-Lautrup type auxiliary variables $B$ and $\bar B$ have been derived from the requirements of nilpotency of order two and absolute anticommutativity of the (anti-)BRST transformations (cf. Eq.~(\ref{36})).

It is evident from the super-expansions of the dynamical variables (see, for e.g. Eqs.~(\ref{23}), (\ref{30}), (\ref{32})) that the Grassmannian translational generators $\big(\frac{\partial}{\partial \theta}\big)\frac{\partial}{\partial \bar \theta}$  provide the geometrical interpretation of the (anti-)BRST transformations $s_{(a)b}$, respectively.
We interpret, on one hand,  that the translation of any generic supervariable $\Phi(t, \theta, \bar \theta)$ along $\bar\theta$-direction keeping $\theta$-direction intact   represents the BRST transformation  of the corresponding generic variable i.e. $\frac{\partial}{\partial \bar \theta}\Phi(t, \theta, \bar \theta) \big|_{\theta = 0}=s_b\,\phi(t)$. On the other hand,  anti-BRST transformation of any generic variable i.e  $s_{ab}\phi(t)$ can be interpreted as the the translation of supervariable along $\theta$-direction keeping  $\bar\theta$-direction fixed i.e. $\frac{\partial}{\partial  \theta}\Phi(t, \theta, \bar \theta) \big|_{\bar \theta = 0} = s_{ab}\phi(t)$. The nilpotency and absolute anticommutativity properties $s_{(a)b}$ have also been captured in terms of Grasmannian translational generators  (cf. Eq.~(\ref{40})).

In view of the present study, we point out that that the current idea can be extended to 
massive/massless spinning charges particle interacting with external electromagnetic field~\cite{mnop} and other models such as Friedmann cosmology~\cite{ren}, closed quantum~\cite{yu} and  supersymmetric (SUSY) cosmological models~\cite{ar,tka}. In latter one, the complex supermultiplets interact with the scalar field. The SUSY cosmological model also exhibits the similar kind of time-reparametrization symmetry and it would be an interesting endeavour to apply the BT superfield/supervariable formalism and the property of (super-)diffeomorphism invariance to obtain the (anti-) BRST symmetry transformations. The BRST quantization of cosmological perturbation using Dirac's prescription has been carried out~\cite{pic}. 
The BT approach within the context of BRST formalism for cosmological perturbation is under investigation and will be reported elsewhere.

\section*{Acknowledgments}
We thankfully acknowledge fruitful suggestions and enlightening comments by our esteemed Referee which have improved the quality of presentation.




\section*{Appendix A: On the derivarion of CF-type of condition}
\label{A1}
In this appendix, we explicitly derive the CF-type of restriction $B+ \bar B + (\dot {\bar C}\,C - \bar C \dot C) = 0$ using the horizontality condition
$\widetilde{\cal N}_M(\widetilde T)\, {\widetilde 
d}\,{\widetilde T}^M = N(t) \,dt$ (cf. Eq.~(\ref{26})). The horizontality condition demands that the Grassmannian components of $\widetilde{\cal N}_M(\widetilde T)\, {\widetilde 
d}\,{\widetilde T}^M $ should be zero. As a result, we obtain the following coefficients of the differential, namely; \\

\noindent
{\textit{Coefficient of $dt:$}}
\begin{eqnarray}
&& \theta \big(\bar \lambda - \bar C \,\dot N - \dot {\bar C} \,N \big) + \bar \theta \big(\lambda - C \,\dot N - \dot C\, N \big) \nonumber\\
&& \hskip 2.5cm +\; \theta \bar \theta \Big(b + \dfrac{d}{dt}\big(h\,N + \bar C\, \lambda - C\, \bar \lambda - \bar C\, C\, \dot N \big) \Big) = 0.
\label{65}
\end{eqnarray}
{\textit{Coefficient of $d \theta:$}}
\begin{eqnarray}
&& \zeta -\bar C \,N + \theta \big(\bar \alpha - \bar C \,\dot \zeta +  \bar C \,\bar \lambda\big) + \bar \theta \big(\alpha + h\, N -  C \,\dot \zeta +\bar C \,\lambda - \bar C \,C \,\dot N \big) \nonumber\\
&& \hskip 2.5cm + \;\theta \bar \theta \big(\sigma + h \,\dot \zeta - h\, \bar \lambda + \bar C\, \dot \alpha - b\, \bar C - C\, \dot{\bar \alpha} + \bar C\, C \,\dot {\bar \lambda}- {\bar C} \,C\, \ddot \zeta \big) =0. 
\label{66}
\end{eqnarray}
{\textit{Coefficient of $d \bar \theta:$}}
\begin{eqnarray}
&& \omega + C\, N + \theta \big(\bar \beta +h \,N - C\, \bar \lambda - \bar C \,\dot \omega - \bar C\, C\, \dot N \big) 
+ \bar \theta \big(\beta - C\, \lambda -  C \,\dot \omega \big) \nonumber\\
&& \hskip 2.5cm + \;\theta \bar \theta \big(\rho + h\, \dot \omega +h \,\lambda +
b\, C - C \,\dot {\bar \beta} + \bar C \,\dot \beta-  \bar C \,C \dot \lambda -\bar C\, C\, \ddot \omega \big) = 0. 
\label{67}
\end{eqnarray}
Comparing the coefficients of Grassmainan coordinates $\theta$, $\bar\theta$ and $\theta \bar\theta$ on the l.h.s. and r.h.s. of Eqs.~(\ref{65}), (\ref{66}) and (\ref{67}), respectively, we obtain the following relationships: 
\begin{eqnarray}
\bar \lambda = \bar C \,\dot N + \dot {\bar C} \,N, \qquad  \lambda = C \,\dot N + \dot C\, N, \qquad  b =-  \dfrac{d}{dt}(h\,N + \bar C\, \lambda - C\, \bar \lambda - \bar C\, C\, \dot N).
\label{68}
\end{eqnarray}
\begin{eqnarray}
&& \zeta = \bar C \,N, \qquad
\bar \alpha = \bar C \,\dot \zeta -  \bar C \,\bar \lambda, \qquad
\alpha = - h\, N +  C \,\dot \zeta -\bar C \,\lambda + \bar C \,C \,\dot N, \nonumber\\
&& \sigma = - h \,\dot \zeta + h\, \bar \lambda - \bar C\, \dot \alpha + b\, \bar C + C\, \dot{\bar \alpha} - \bar C\, C \,\dot {\bar \lambda}+ {\bar C} \,C\, \ddot \zeta.
\label{69}
\end{eqnarray}
\begin{eqnarray}
&& \omega =- C\, N, \qquad
\bar \beta = -h \,N + C\, \bar \lambda + \bar C \,\dot \omega + \bar C\, C\, \dot N, \qquad 
\beta = C\, \lambda +  C \,\dot \omega, \nonumber\\
&& \rho =- h\, \dot \omega -h \,\lambda -
b\, C + C \,\dot {\bar \beta} - \bar C \,\dot \beta +  \bar C \,C \dot \lambda +\bar C\, C\, \ddot \omega.
\label{70}
\end{eqnarray}
Exploiting these equations, we obtain the various relationships as listed in Eq.~(\ref{29}) amongst the various secondary variables and basic (dynamical) variables of the (anti-)BRST invariant theory. Subtracting second relation of Eq.~(\ref{70}) from the third relation of Eq.~(\ref{69}) and using the values of $\zeta, \, \bar\lambda,\, \omega$ and $\lambda$,  we obtain 
\begin{eqnarray}
\alpha - \bar \beta = C(\dot \zeta - \bar \lambda) - \bar C(\dot \omega +\lambda) = 0 \Rightarrow \alpha = \bar \beta. 
\label{71}
\end{eqnarray}  
Similarly, one can also prove other relationships in Eq.~(\ref{29}). Now we make the choices (without any loss of generality) for the Nakanishi-Lautrup type auxiliary variables $B$ and $\bar B$ as $B = - (h + \bar C \,\dot C)$ and $\bar B = +(h + \bar C \, \dot C)$. As a result, the expressions of secondary variables $\alpha$ and $\beta$ read (cf. Eq.~(\ref{29}))
\begin{eqnarray}
\alpha = N\,B - \bar C \big(\dot C\, N + C \,\dot N \big),  \qquad
\bar \beta = - N \,\bar B - \big(\dot  {\bar C}\, N  + \bar C \,\dot N \big) C.
\label{72}
\end{eqnarray}
Substituting the values of $\alpha$ and $\beta$ in Eq.~(\ref{71}), we obtain the celebrated CF-type of condition $B+ \bar B + (\dot {\bar C}\,C - \bar C \dot C) = 0$.

\end{document}